\documentstyle[12pt]{article}
\begin{document}

\begin{titlepage}
\baselineskip 20pt

\title{The box anomaly and radiative decays of $\eta(\eta^\prime$)-mesons}

\vspace{0.5cm}

\author{M. A. Ivanov\thanks{Permanent address: Bogoliubov Laboratory of
Theoretical Physics, Joint Institute for Nuclear Research, 141980 Dubna
(Moscow Region), Russia}
\, and T. Mizutani\\
\\
\\
Department of Physics\\
Virginia Polytechnic Institute and State University\\
Blacksburg, VA 24061}
\maketitle

\abstract
\baselineskip 20pt

We report on a confined quark model calculation of the transition form factors
for the radiative decays of $\eta(\eta^\prime)\to\gamma l^+l^-$
and $\eta(\eta^\prime)\to\gamma\pi^+\pi^-$.
The $\eta/\eta^\prime$ mixing angle $\theta_P$
is determined from the set of data on the electromagnetic $\eta(\eta^\prime)$
decays. It is found that $\theta_P\sim -16.5^{\rm o}$. 

The analysis of the dipionic mass distribution in the decay
$\eta^\prime\to\gamma\pi^+\pi^-$ confirms the existence
of a non-resonant contribution which is the box anomaly.

\bigskip
\noindent 
PACS number(s): 12.39.-x,12.39.Ki,12.40.Vv,11.30.Rd,
13.40.Gp,13.40.Hq,14.40.-n.

\end{titlepage}

\section{Introduction}
\baselineskip 20pt

The interest of studying the properties of the neutral pseudoscalar particles
$\pi^0$, $\eta$ and $\eta^\prime$ is not weaken from the begining of
creation of unitary SU(3) symmetry. One of the intrigued problem  was 
the so-called "U(1)-problem" which was originally encountered by Glashow
\cite{Glashow}. He considered extending SU(3)$\times$ SU(3) to
U(3)$\times$ U(3) and found that if the extra U(1) axial symmetry 
was realized spontaneously, there would have to be an extra light
pseudoscalar meson with mass compatible with pion (140 MeV). But the only
flavor singlet meson is $\eta^\prime$ with a mass being equal to 958 MeV!
After the advent of QCD, one could suggest that the anomaly associated with
axial U(1) transformations might be used to provide a solution
of U(1)-problem because of the divergence of axial current does not vanish
in the chiral limit. However, the matter was found to be not so simple
(see, for review \cite{Christos}). Witten \cite{Witten} found that
the anomaly could be turned off as a number of quark colors goes to infinity
($N_c\to\infty$) and then the $\eta^\prime$-meson would be considered
as the Nambu-Goldstone boson. With finite $N_c$ he suggested that the mass
squared of the $\eta^\prime$ was proportional to $1/N_c$. Following
this way, Veneziano \cite{Venez}  proposed a more explicit mechanism
which was consistent with anomalous Ward identities and realized it
by using the effective chiral Lagrangian incorporated explicitly $U_A(1)$
anomaly. The phenomenological implications of those Lagrangians
to the $\eta-\eta^\prime$-mixing allowed one to get the mass formulas
for the masses of $\eta$ and $\eta^\prime$ and the mixing angle
which reproduce the experimental values with quite resonable accuracy.

The three-flavor Nambu-Jona-Lasinio (NJL) model with the instanton-induced
six-quark effective interaction which incorporates effects of $U_A(1)$ anomaly
has been developed to study the spectrum of low-lying mesons, 
the $\eta-\eta^\prime$-mixing, etc. \cite{NJL}.

The lattice calculations of the $\eta^\prime$ mass have been performed
in \cite{LAT} with taking into account U(1) gluon contributions
which confirm that the $\eta^\prime$ owes its mass to the topological
susceptibility.

One can say that the investigation of the internal structure of
the neutral pseudoscalar (also vector) mesons is of great interest.
As well known that very good probe of hadronic structure is the photon,
for example, the study of electromagnetic form factors gives us
the knowledge on the internal structure of charge particles.
However, the geniune neutral particles like $\pi^0$, $\eta$, $\eta^\prime$
mesons have no the electromagnetic form factors because of C-invariance.
Therefore, it is needed to study the two-photon processes like
$P\to\gamma\gamma$ ($P=\pi^0$, $\eta$, $\eta^\prime$) with one
or both photons being off mass shell to probe their structure.

One has to remark that theoretically the decay $\pi^0\to\gamma\gamma$
with both of photons being on mass shell is defined by the Adler-Bell-Jackiw
(ABJ) triangle anomaly whereas the gluon contribution from $U_A(1)$ anomaly
should be taken into account under derivation of expression for
$\eta^\prime(\eta)\gamma\gamma$ coupling \cite{Shore}. In many papers
people do not use a such complication and just employ the parametrization
of those amplitudes in terms of the decay constants $F_\pi$, $F_\eta$,
$F_{\eta^\prime}$, and the $\eta-\eta^\prime$ mixing angle which are
determined by fitting the relevant experimental data \cite{Gilman}.
A precise relation has been established between these phenomenological
parameters and those of the chiral expansion \cite{Mouss}.

The more deep insight to the structure of neutral particle may be done
by looking at the photon-dilepton decays $P\to\gamma l^+l^-$
(and similar decays of the neutral vector mesons $V\to P l^+l^-$ 
with V being $\rho^0$, $\omega$, $\phi$) (see, review \cite{Land}). 
These processes are determined by the transition form factor $F_{PVV}(q^2)$
in the time-like region of momentum transfer squared. Obviously, it is not
enough to use the effective chiral lagrangians to obtain such form factor.
It is needed to involve some assumptions on dynamics of such transitions
like the vector dominance model (VDM) or quark models.
The transition form factors in space-like region ($Q^2=-q^2\ge 0$)
have been studied experimentally in the reaction $e^+e^-\to e^+e^-P$
and are planning to study in the photo and electroproduction of $\eta^\prime$
from nucleons and nuclei on CEBAF's energy using the "virtual" Primakoff effect
\cite{David}.

Another processes which help to shed light on the structure of the 
$\eta(\eta^\prime)$ mesons are decays $\eta(\eta^\prime)\to\pi^+\pi^-\gamma$
which are described theoretically by both the box-anomaly and 
resonance contribution.. The phenomenolgical
analysis of these processes have been done in paper \cite{Benay}.

To summarize this brief sketch, we would like to emphasize once more that 
many efforts have been
devoted to describing the low-energy hadron physics in terms of effective 
chiral Lagrangians which include the Wess-Zumino anomaly 
\cite{Gasser}-\cite{Hols}. There low-energy QCD is
described in terms of mesonic degrees of freedom instead of those quarks and
gluons with  taking into account the underlying symmetries and anomalies of
QCD.  The hadronic verteces in the effective Lagrangians are structureless
therefore the simple way to take into account the momentum dependence
has been chosen by incorporating the vector mesons either as massive
Yang-Mills bosons or as the dynamical gauge bosons of a hidden local symmetry
in the nonlinear chiral Lagrangians \cite{Meiss}, \cite{Picc}.

We also note that the popular methods of QCD sum rules and lattice
QCD are limited for timelike region. There are a few QCD-motivated
quark models which incorporate quark degrees of freedom and allow one
to describe many-particle processes both at spacelike and timelike regions.
One of them \cite{Craig} is based on the QCD Dyson-Schwinger equations with
solution describing a confined quark, and another one \cite{EI}, \cite{Ivmiz} is
based on the phenomenological treatment of  the mesons as $q\bar q$ states
with  the confinement ansatz to ensure the absence of quark production
thresholds  in the S-matrix elements of physical processes. Both of them have
been successfully applied to describing the low-energy processes such as
the reaction  $\gamma\pi\to\pi\pi$.
The salient feature of our confined quark approach \cite{EI},
\cite{Ivmiz} is that all mesons
(pseudoscalar, vector, scalar, etc.) are described as $q\bar q$ states
by writing down the interaction Lagrangians and employing the
compositeness condition to determine the meson-quark vertex coupling constants.
The S-matrix elements are calculated in the lower order in the $1/N_c$ expansion
which automatically includes the diagrams with intermediate resonances.

In the present article we report on a calculation of the transition form
factors of the processes $P\to\gamma l^+l^-$ and $V\to Pl^+l^-$ 
which are related to the study of momentum dependence of triangle anomaly, and 
the $\eta(\eta^\prime)\to\gamma\pi^+\pi^-$-decays which are related to
the box anomaly  within a confined quark model. This is a continuation
of study the momentum dependence of anomalous processes which we started
in the paper \cite{IvanMiz}.

\section{Model}
\baselineskip 20pt

A confined quark model is specified by the interaction Lagrangian
describing the transition mesons into quarks

\begin{equation}\label{lagran}
L_I(x)=
ig_P \bar q(x)P(x)\gamma^5 q(x)+
g_V  \bar q(x)V^\mu(x)\gamma^\mu q(x)+
eA^\mu(x)\bar q(x)Q\gamma^\mu q(x)
\end{equation}
with the lowest lying pseudoscalar and vector nonet meson matrices
being equal to

\[
P=\left( \begin{array}{ccc}
\mbox{$\sqrt{\frac{1}{2}}\pi^0+
       \sqrt{\frac{1}{6}}\eta_8+
       \sqrt{\frac{1}{3}}\eta_1 $} &
\mbox{$\pi^+$} & \mbox{$K^+$} \\
\mbox{$\pi^-$} &
\mbox{$ -\sqrt{\frac{1}{2}}\pi^0+
         \sqrt{\frac{1}{6}}\eta_8+
         \sqrt{\frac{1}{3}}\eta_1$} &  \mbox{$K^0$} \\
\mbox{$K^-$} & \mbox{$\bar K^0$} &
\mbox{$ -\sqrt{\frac{2}{3}}\eta_8+\sqrt{\frac{1}{3}}\eta_1 $}
\end{array} \right)
\]

\[
V=\left( \begin{array}{ccc}
\mbox{$ \sqrt{\frac{1}{2}}\rho^0+
        \sqrt{\frac{1}{6}}\omega_8+
        \sqrt{\frac{1}{3}}\omega_1 $} & \mbox{$\rho^+$} & \mbox{$K^{*+}$} \\
\mbox{$\rho^-$} &
\mbox{$ -\sqrt{\frac{1}{2}}\rho^0+
         \sqrt{\frac{1}{6}}\omega_8+
         \sqrt{\frac{1}{3}}\omega_1 $} & \mbox{$K^{*0}$} \\
\mbox{$K^{*-}$} & \mbox{$\bar K^{*0}$} &
\mbox{$-\sqrt{\frac{2}{3}}\omega_8+\sqrt{\frac{1}{3}}\omega_1$}
\end{array} \right)
\]
The physical fields $\eta$, $\eta^\prime$, and $\omega$, $\phi$ are defined
in the standard manner by introducing the mixing angles as

\begin{eqnarray}\label{mix}
\eta&=&-\eta_1\sin{\theta_P}+\eta_8\cos{\theta_P} \hspace{1cm}
\eta^\prime=\eta_1\cos{\theta_P}+\eta_8\sin{\theta_P} \nonumber\\
&&\\
\phi&=&-\omega_1\sin{\theta_V}+\omega_8\cos{\theta_V} \hspace{1cm}
\omega=\omega_1\cos{\theta_V}+\omega_8\sin{\theta_V} \nonumber
\end{eqnarray}

so that their interaction with quarks may be written as

\begin{eqnarray}
L_I^{\eta,\eta^\prime}&=&
-ig_\eta\eta \{
\frac{\bar u\gamma^5 u+\bar d\gamma^5 d}{\sqrt{2}}\sin{\delta_P}+
\bar s\gamma^5 s \cos{\delta_P} \} \nonumber\\
&& \label{etamix} \\
&&+ig_{\eta^\prime}\eta^\prime \{
\frac{\bar u\gamma^5 u+\bar d\gamma^5 d}{\sqrt{2}}\cos{\delta_P}-
\bar s\gamma^5 s \sin{\delta_P} \} \nonumber\\
&&\nonumber\\
&&\nonumber\\
L_I^{\omega,\phi}&=&
-g_\phi\phi^\mu \{
\frac{\bar u\gamma^\mu u+\bar d\gamma^\mu d}{\sqrt{2}}\sin{\delta_V}+
\bar s\gamma^\mu s \cos{\delta_V} \} \nonumber\\
&& \label{vectormix} \\ 
&&+ig_{\omega}\omega^\mu \{
\frac{\bar u\gamma^\mu u+\bar d\gamma^\mu d}{\sqrt{2}}\cos{\delta_V}-
\bar s\gamma^\mu s \sin{\delta_V} \} \nonumber
\end{eqnarray}
where $\delta=\theta-\theta_I$. The "ideal" mixing
angle is equal to $\theta_I=\arctan({1/\sqrt{2}})=35.3^{\rm o}$. 

The mixing angles are usually determined from the mass formular (quadratic or 
linear) \cite{PDG}:
$\theta^{\rm quad}_P=-11^{\rm o}$, $\theta^{\rm lin}_P=-23^{\rm o}$,
and  $\theta^{\rm quad}_V=39^{\rm o}$, $\theta^{\rm lin}_V=36^{\rm o}$.
This means that $\theta_V$ is very close to $\theta_I$, i.e. 
$\phi\approx s\bar s$ and we will use this value in what follows.

The Veneziano's model \cite{Venez} which incorporates the expicit $U_A(1)$
anomaly gives $\theta_P=-18^{\rm o}$.

In our model we use the experimental values of all hadron masses under
calculations of matrix elements and consider the mixing angles defined
by Eq.~(\ref{mix}) as the adjustable parameters. 
We determine $\theta_P$ by fitting the available experimental data.

The meson-quark coupling constants $g_M$ are defined by the 
$compositeness \, \, \, condition$ meaning  that the renormalization
constants of the meson fields are equal to zero

\begin{equation}\label{z=0}
Z_M=1-g^2_M\Pi_M^\prime(m^2_M)=0.
\end{equation}
Here $\Pi_M^\prime$ is the derivative of the meson mass operator and
$m_M$ is the physical meson mass.
In other words, the equation (\ref{z=0}) provides the right normalization
of the charge form factor $F_M(0)=1$. This could be readily seen from
the Ward identity

$$
g^2_M\Pi_M^\prime(p^2)=g^2_M{1\over 2p^2}p^\mu
{\partial \Pi_M(p^2)\over\partial p^\mu}=g^2_M{1\over 2p^2}p^\mu
T^\mu_M(p,p)=F_M(0)=1.
$$
where $T_M^\mu(p,p^\prime)$ is the three-point function  describing charge
form factor.

Mesonic interactions in the QCM are defined by the closed quark loops

\begin{equation}\label{loop}
\int\!\! d\sigma_v {\rm tr}\biggl[M(x_1)S_v(x_1-x_2)...M(x_n)S_v(x_n-x_1)
\biggr].
\end{equation}
Here,

\begin{equation}\label{light}
S_v(x_1-x_2)=\int\!\!{d^4p\over (2\pi)^4i} e^{-ip(x_1-x_2)}
{1\over v\Lambda-\not\! p}
\end{equation}
is a quark propagator with the scale parameter $\Lambda$ characterizing
the size of the confinement, and the measure $d\sigma_v$, which is essential
for quark confinement, is defined to provide the absence of singularities
in Eq.~(\ref{loop}) corresponding to the physical quark production:

\begin{equation}\label{conf}
\int\!\!{d\sigma_v\over v-z}\equiv G(z)=a(-z^2)+zb(-z^2).
\end{equation}

The shapes of the confinement functions $a(u)$ and $b(u)$, and the scale
parameter $\Lambda$ have been determined from the best model description
of data in low-energy processes:

\begin{equation}\label{confunc}
a(u)=2\exp(-u^2-u) \hspace{1cm} b(u)=2\exp(-u^2+0.4u) \hspace{1cm}
\Lambda=460 {\rm MeV},
\end{equation}
which describe various basic constants quite well (see, \cite{EI}, 
\cite{Ivmiz}).

The analytical expressions and numerical values for the meson-quark
coupling constants from Eq.~(\ref{z=0}) are written down

\[
g_P=2\pi\sqrt{\frac{2}{3R_{PP}(\mu_P^2)}}=\left\{ \begin{array}{ll}
 3.37 & \mbox{$\pi$} \\
 2.92 & \mbox{$\eta$} \\
 2.65 & \mbox{$\eta^\prime$} 
\end{array} \right.
\]

\[
g_V=2\pi\sqrt{\frac{1}{R_{VV}(\mu_V^2)}}=\left\{ \begin{array}{ll}
 3.24 & \mbox{$\rho$} \\
 3.23 & \mbox{$\omega$} \\
 3.17 & \mbox{$\phi$} 
\end{array} \right.
\]

We use the notation $\mu=m/\Lambda$ throughout. The structural integrals 
$R_{PP}(x)$ and $R_{VV}(x)$ are given in the Table I.

Also we have the possibility to investigate the consequences of 
the explicit $U_A(1)$ breaking in our approach. To do it, we may just assume
that the octet and singlet coupling constants in the Lagrangian (\ref{lagran})
are not equal each other

\begin{equation}
L_{81}=\frac{i}{\sqrt{2}}\bar q \biggl(g_8\eta_8\lambda^8+g_1\eta_1\lambda^0
\biggr)\gamma^5 q
\end{equation}
where $\lambda^0=\sqrt{2/3} I$. Using this Lagrangian in the compositeness 
condition (\ref{z=0}) and defining the effective couplings
of physical fields with quarks give

\begin{eqnarray}
g^2_\eta &\equiv & g^2_8 \cos^2\theta+g^2_1 \sin^2\theta=
\frac{1}{\Pi^\prime_P(m^2_\eta)}\nonumber\\
&&\label{g8g1}\\
g^2_{\eta^\prime} &\equiv & g^2_8 \sin^2\theta+g^2_1 \cos^2\theta=
\frac{1}{\Pi^\prime_P(m^2_{\eta^\prime})}.
\nonumber
\end{eqnarray}

Then we introduce two new angles characterizing the $\eta-\eta^\prime$-
mixing with taking into account the $U_A(1)$-breaking:

$$
\sin\theta_\eta=\frac{g_1}{g_\eta} \sin\theta \hspace{1cm}
\sin\theta_{\eta^\prime}=\frac{g_8}{g_\eta} \sin\theta.
$$
Thus the explicit $U_A(1)$ breaking  gave us two different mixing angles
which are determined by the $\eta$ and $\eta^\prime$ masses through
the compositeness condition and the initial mixing angle $\theta$.
Their numerical values are not much differ from this angle, for example,
if $\theta=-11^{\rm o}$ then $\theta_\eta=-9.9^{\rm o}$ and 
$\theta_{\eta^\prime}=-12.2^{\rm o}$ , 
if $\theta=-18^{\rm o}$ then $\theta_\eta=-16^{\rm o}$ and 
$\theta_{\eta^\prime}=-20.2^{\rm o}$. The available
experimental data on the $\eta(\eta^\prime)$-decays have too large uncertainties
to distinguish these possibilities. So we will not consider the explicit
$U_A(1)$-breaking for the time being.

\section{Radiative decays of neutral pseudoscalar and 
vector mesons}

\subsection{$P\to\gamma l^+l^-$ and $V\to P l^+l^-$ decays}
\baselineskip 20pt

The study of electromagnetic structure of neutral mesons
($P=\pi^0$, $\eta$, $\eta^\prime$, $V=\rho^0$, $\omega$, $\phi$) may be done
in the two-photon processes: $P\to\gamma\gamma$, $P\to\gamma l^+l^-$,
$P\to l^+ l^-$, or in the processes  with a change of C-parity of the mesons
in the initial and final states: $V\to P l^+l^-$ and $P\to V l^+ l^-$.

First, we give the necessary model-independent formulas 
(invariant matrix elements and  differential distributions)
which we use in our calculations.

\vspace{1cm}
\noindent
{\bf (a) $P\to \gamma l^+l^-$. }

\begin{eqnarray}
M(P\to\gamma l^+l^-)&=&e^3G_{P\gamma\gamma}(m_P^2,0,q^2)
\varepsilon^{\mu\nu\alpha_1\alpha_2}\epsilon^\nu q_1^{\alpha_1} q^{\alpha_2}
\frac{1}{q^2} \bar l\gamma^\mu l \nonumber\\
&&\nonumber\\
\frac{d\Gamma(P\to\gamma l^+l^-)} {dq^2\cdot \Gamma(P\to\gamma\gamma)}&=&
\frac{2\alpha}{3\pi} \frac{1}{q^2}
                           \sqrt{1-\frac{4m^2_l}{q^2}}
                                   \biggl[1+\frac{2m^2_l}{q^2}\biggr]
 \biggl[1-\frac{q^2}{m^2_P}\biggr]^3 |F_P(q^2)|^2
\label{spec_p}\\
&&\nonumber\\
F_P(q^2)&=&G_{P\gamma\gamma}(m_P^2,0,q^2)/G_{P\gamma\gamma}(m_P^2,0,0)
\nonumber\\
&&\nonumber\\
\Gamma(P\to\gamma\gamma)&=&
\frac{\alpha^2\pi}{4}m_P^3 G^2_{P\gamma\gamma}(m^2_P,0,0),
\nonumber
\end{eqnarray}
Here $p$ ($p^2=m_P^2$), $q_1$ ($q_1^2=0$), and $q$ are the four-momentum
of pseudoscalar, photon, and lepton pair, respectively, and $\epsilon^\nu$ is
the polarization vector of outgoing photon.

\vspace{1cm}

\noindent
{\bf (b) $V\to P l^+l^-.$ }

\begin{eqnarray}
M(V\to P l^+l^-)&=&e^2G_{VP\gamma}(m_V^2,m_P^2,q^2)
\varepsilon^{\mu\nu\alpha_1\alpha_2}\epsilon^\nu q_1^{\alpha_1} q^{\alpha_2}
\frac{1}{q^2} \bar l\gamma^\mu l \nonumber\\
&&\nonumber\\
\frac{d\Gamma(V\to P l^+l^-)} {dq^2\cdot \Gamma(V\to\gamma P)}&=&
\frac{\alpha}{3\pi} \frac{1}{q^2}
                           \sqrt{1-\frac{4m^2_l}{q^2}}
                                   \biggl[1+\frac{2m^2_l}{q^2}\biggr] \cdot
\label{spec_v}\\
&& \biggl[1-\frac{q^2}{(m_V-m_P)^2}\biggr]^{3/2}
 \biggl[1-\frac{q^2}{(m_V+m_P)^2}\biggr]^{3/2}
|F_{VP}(q^2)|^2
\nonumber\\
&&\nonumber\\
F_{VP}(q^2)&=&
G_{VP\gamma}(m_V^2,m_P^2,q^2)/G_{VP\gamma}(m_V^2,m_P^2,0)
\nonumber\\
&&\nonumber\\
\Gamma(V\to P\gamma)&=&\frac{\alpha}{24}m_V^3 
\biggl[1-\frac{m^2_P}{m^2_V}\biggr]^3 G^2_{VP\gamma}(m^2_V,m^2_P,0)
\nonumber
\end{eqnarray}
Here $p$ ($p^2=m_V^2$), $q_1$ ($q_1^2=m_P^2$), and $q$ are 
the four-momentum of vector, pseudoscalar meson, and lepton pair, respectively,
$\epsilon^\nu$ is the polarization vector of incoming vector meson.

The transition form factors $F_P(q^2)$ and $F_{VP}(q^2)$ define the internal
structure of the neutral mesons and should be calculated dynamically.

In our approach they are defined by the diagrams in Fig. 1 for the decay
$P\to\gamma l^+l^-$ and similar diagrams for the decay $V\to P l^+l^-$.

We have

\begin{eqnarray}
G_{P\gamma\gamma}(p^2,q_1^2,q^2)&=& g_{P\gamma\gamma}(p^2,q_1^2,q^2)+
q^2\sum\limits_{V}g_{VP\gamma}(p^2,q_1^2,q^2) g_{V\gamma}(q^2) D_V(q^2)
\label{PGG}\\
&&\nonumber\\
G_{VP\gamma}(p^2,q_1^2,q^2)&=&g_{VP\gamma} (p^2,q_1^2,q^2)+
q^2 \sum\limits_{V^\prime}  g_{VPV^\prime} (p^2,q_1^2,q^2)
g_{V^\prime\gamma} (q^2) D_{V^\prime}(q^2)
\label{VPG}
\end{eqnarray}
Here the first term corresponds to the triangle diagram whereas
the second to the vector meson pole diagram in Fig. 1. One should emphasize
that in our approach the contribution from the pole diagram vanishes if 
the photon is on mass shell ($q^2=0$) as a consequence of gauge invariance.

The quantities appearing Eqs.~(\ref{PGG}-\ref{VPG}) are written down 
 
\begin{eqnarray}
g_{P\gamma\gamma}(p^2,q_1^2,q_2^2)&=&\frac{g_P}{2\sqrt{2} \pi^2}
\frac{1}{\Lambda}C_{P\gamma\gamma}
R_{PVV}(p^2/\Lambda^2,q_1^2/\Lambda^2,q_2^2/\Lambda^2)
\nonumber\\
& & \nonumber\\
g_{VP\gamma}(p^2,q_1^2,q_2^2)&=&\frac{3g_P g_V}{8\pi^2}\frac{1}{\Lambda}
C_{VP\gamma}
R_{PVV}(p^2/\Lambda^2,q_1^2/\Lambda^2,q_2^2/\Lambda^2)
\nonumber\\
& & \nonumber\\
g_{V\gamma}(q^2)&=&\frac{g_V}{4\sqrt{2}\pi^2}C_{V\gamma}
R_V(q^2/\Lambda^2)
\nonumber
\end{eqnarray}

We do not write  the expressions for $g_{VPV^\prime}$ because of we will 
consider the only decay $\omega\to\pi^0 \mu^+ \mu^-$. In this case 
$g_{\omega\pi^0\rho}=(3 g_\pi g_\rho g_\omega/2\sqrt{2}\pi^2) 
R_{PVV}(p^2/\Lambda^2,q_1^2/\Lambda^2,q_2^2/\Lambda^2)$.

The structural integrals and SU(3)-factors entering these expressions
are given in Table I and II.

The Eqs.~(\ref{PGG}-\ref{VPG}) may be written in the forms which are more 
convenient for comparison with the VDM approach:

\begin{eqnarray}
G_{\pi^0\gamma\gamma}(p^2,q_1^2,q^2)&=&
g_{\pi^0\gamma\gamma}(p^2,q_1^2,q^2)
\biggr\{
1+q^2\biggr[\frac{1}{2} C_\rho(q^2)+\frac{1}{2} C_\omega(q^2)\biggl]
\biggl\}
\nonumber\\
&&\nonumber\\
G_{\eta\gamma\gamma}(p^2,q_1^2,q^2)&=&
g_{\eta\gamma\gamma}(p^2,q_1^2,q^2)\cdot 
\biggr\{
1+q^2\frac{3}{5\sin\delta_P+\sqrt{2}\cos\delta_P}\cdot\biggl.
\nonumber\\
&&\biggr.
\biggr[\frac{3\sin\delta_P}{2} C_\rho(q^2)+
       \frac{\sin\delta_P}{6} C_\omega(q^2)+
       \frac{\sqrt{2}\cos\delta_P}{3} C_\phi(q^2)\biggl]
\biggl\}
\nonumber\\
&&\nonumber\\
G_{\eta^\prime\gamma\gamma}(p^2,q_1^2,q^2)&=&
g_{\eta^\prime\gamma\gamma}(p^2,q_1^2,q^2)\cdot 
\biggr\{1+q^2 \frac{3}{5\cos\delta_P-\sqrt{2}\sin\delta_P}\cdot \biggl.
\nonumber\\
&&\biggr.
\biggr[\frac{3\cos\delta_P}{2} C_\rho(q^2)+
       \frac{\cos\delta_P}{6} C_\omega(q^2)-
       \frac{\sqrt{2}\sin\delta_P}{3} C_\phi(q^2)\biggl]
\biggl\}
\nonumber
\end{eqnarray}
where

\begin{equation}\label{coef}
C_V(q^2)=\frac{g^2_V}{4\pi}D_V(q^2)R_V(q^2/\Lambda^2)=
\frac{R_V(q^2/\Lambda^2)}{R_{VV}(m^2_V/\Lambda^2)}D_V(q^2)
\end{equation}
with $D_V$ being a vector meson propagator. 

The expression for the renormalized vector propagator obtained within our 
model  \cite{ijmp} in one-loop approximation is written 

\begin{equation}\label{prop_QCM}
D_V(q^2)=\frac{R_{VV}(m^2_V/\Lambda^2)}
{m^2_V R_V(m^2_V/\Lambda^2)-q^2 R_V(q^2/\Lambda^2)}
\end{equation}
where the obvious equality (see, Table 1) $ (xR_V(x))^\prime=R_{VV}(x)$
provides the residue at $q^2=m^2_V$ to be equal to one.

The imaginary part of the $\rho(\omega)$-meson propagators is not very 
important for  $\pi$ and $\eta$ decays but it should be taken into account 
somehow for $\eta^\prime$ decay. In our approach the imaginary part should
appear as a result of summing up the two-loop diagrams. It will be proportional
to the whole decay width of a vector meson, for example, for $\rho$-meson
it reads $\Gamma_\rho(q^2)=m_\rho (g_{\rho\pi\pi}^2(q^2)/48\pi)
[1-4m_\pi^2/q^2]^{3/2}\Theta(q^2-4m_\pi^2) $  where
$g_{\rho\pi\pi}(q^2)=(3g_\pi^2 g_\rho / 4\sqrt{2}\pi^2) R_{VPP}(q^2)$.

Then the vector propagator may be written as

\begin{equation}\label{propim_QCM}
D_V(q^2)=\frac{R_{VV}(m^2_V/\Lambda^2)}
{m^2_V R_V(m^2_V/\Lambda^2)-q^2 R_V(q^2/\Lambda^2)-im_V \Gamma_V R_{VV}(m^2_V)}
\end{equation}
with $\Gamma_V$ being the experimental value for vector meson decay. We drop
out the $q^2$ dependence of decay width since it is the resonance region.
The extra factor $R_{VV}(m^2_V)$ in denominator is introduced to provide the 
correct normalization of propagator at $q^2=m^2_V$.
Also we will use the same magnitude of width for $\rho$ and $\omega$
mesons $\Gamma_V=151$ MeV because of $\omega$ has too small width (about 8 MeV)
to be seen in the experiments and its account theoretically will just give
unpleasant extra pick. Of course, one drops $\phi$-width because
it is not important at energy less 1 GeV.

For comparison, we will use the simplified VDM-form for vector propagators

\begin{equation}\label{propim_VDM}
D_V(q^2)=\frac{1}{m^2_V-q^2 -i m_V \Gamma_V}.
\end{equation}

 We will discuss the numerical results for decay widths and the form
factors in the final subsection.
 
\subsection{$\eta(\eta^\prime)\to\pi^+\pi^-\gamma$ decay}
\baselineskip 20pt

First, we give the model-independent expessions for invariant matrix element
decay width and distributions of the $P\to\pi^+\pi^-\gamma$-decay
(P=$\eta$ or $\eta^\prime$):

\begin{eqnarray}
M^\mu(P\to\pi^+\pi^-\gamma)&=&
e \varepsilon^{\mu\alpha\beta\nu}p_+^\alpha p_-^\beta q^\nu
G_{P\to\pi^+\pi^-\gamma}(s_1,s_2,s_3)\nonumber\\
&&\label{EPPG}\\
\Gamma(P\to\pi^+\pi^-\gamma)&=&
\frac{\alpha}{256\pi^2} \frac{1}{m_P^3}
\int\limits_{4m^2_\pi}^{m^2_P}ds_2 s_2
\int\limits_{s_1^-}^{s_1^+}ds_1(s_1^+-s_1)(s_1-s_1^-)
|G_{P\to \pi^+\pi^-\gamma}(s_1,s_2,s_3)|^2 
\nonumber
\end{eqnarray}
Here, the kinematical variables $s_i$ are defined as
$$
s_1=(q+p_+)^2 \hspace{0.5cm} s_2=(p_-+p_+)^2 \hspace{0.5cm}
s_3=(q+p_-)^2, \hspace{0.5cm} s_1+s_2+s_3=m^2_P+2m^2_\pi.
$$
so that they vary in the intervals $4m^2_\pi\le s_2\le m^2_P$ and
$ s_1^-\le s_1 \le s_1^+$ where \\
$s_1^{\pm}=m^2_\pi+(1/2)(m^2_P-s_2)[1\pm \sqrt{1-4m^2_\pi/s_2}]$.

In our approach the matrix element of the decay $P\to \pi^+\pi^-\gamma$ 
is defined by the diagrams in Fig. 2. 
The expression for  $G_{P\to \pi^+\pi^-\gamma}$ is written down

\begin{eqnarray}
G_{P\to\pi^+\pi^-\gamma}(s_1,s_2,s_3)&=&
g_{P\to\pi^+\pi^-\gamma}^{\rm box}(s_1,s_2,s_3)+
2g_{P\rho\gamma}(s_2)g_{\rho\pi\pi}(s_2)D_\rho(s_2)\nonumber\\
&&\nonumber\\
g_{P\rho\gamma}(s_2)&=&C_P\frac{1}{\Lambda}
\frac{3g_P g_\rho}{4\pi^2}
R_{PVV}(m^2_P/\Lambda^2,s_2/\Lambda^2,0)
\nonumber\\
&&\nonumber\\
g_{P\to\pi^+\pi^-\gamma}^{\rm box}(s_1,s_2,s_3)&=&
C_P\frac{1}{\Lambda^3}\frac{3g^2_\pi g_P}{4\pi^2\sqrt{2}}
R_{\Box}(s_1/\Lambda^2,s_2/\Lambda^2,s_3/\Lambda^2)
\nonumber\\
&&\nonumber\\
R_{\Box}(s_1,s_2,s_3)&=&
\int d^4\alpha\delta(1-\sum\limits_{i=1}^{4}\alpha_i)
\{
-a^\prime(-D_4^{-+\gamma})+a^\prime(-D_4^{+\gamma-})
-a^\prime(-D_4^{\gamma-+})
\}
\nonumber
\end{eqnarray}
where
\begin{eqnarray}
D_4^{-+\gamma}&=&\alpha_3\alpha_4 m^2_P+
\alpha_2(\alpha_1+\alpha_3)m^2_\pi
+\alpha_1\alpha_3 s_2+\alpha_2\alpha_4 s_1
\nonumber\\
D_4^{+\gamma-}&=&\alpha_2\alpha_3 m^2_P+
(\alpha_3\alpha_4+\alpha_1\alpha_2)m^2_\pi
+\alpha_1\alpha_3 s_3+\alpha_2\alpha_4 s_1
\nonumber\\
D_4^{\gamma-+}&=&\alpha_1\alpha_2 m^2_P+
\alpha_3(\alpha_2+\alpha_4)m^2_\pi
+\alpha_1\alpha_3 s_3+\alpha_2\alpha_4 s_2
\nonumber
\end{eqnarray}
The group coefficient $C_P$ is equal to
$-\sin\delta_P$ for $\eta$ and $\cos\delta_P$ for $\eta^\prime$.

One trivially  finds that $R_\Box(0,0,0)=1/3$. In the physical
region the box-contribution varies very slowly around this value so that we 
may approximate $R_\Box$ by this value with quite good accuracy 
in the calculations.  

This gives the following formula for the decay width

\begin{equation}\label{etawidth}
\Gamma(P\to\pi^+\pi^-\gamma)=
\frac{\alpha}{12\pi^2} 
\int\limits_{2m_\pi}^{m_P} dm E_\gamma^3 P_\pi^3
|G^{(0)}_{P\to\pi^+\pi^-\gamma}(m^2)|^2.
\end{equation}
with $E_\gamma=(m^2_P-m^2)/2m_P$ and $P_\pi=(1/2)\sqrt{m^2-4m_\pi^2}$
being  the photon energy in the ingoing meson rest system and
the pion momentum in the dipion rest system, respectively. 
The label (0) upstairs  $G$ means that we use $R_\Box=1/3$
in the whole physical region.

The effective mass spectrum of the $\pi^+\pi^-$ system is defined as

\begin{equation}\label{masspectr}
\frac{d\Gamma}{dm_{\pi\pi}}=
\frac{\alpha}{12\pi^2} E_\gamma^3 P_\pi^3
|G^{(0)}_{\eta\pi^+\pi^-\gamma}(m^2_{\pi\pi})|^2.
\end{equation}

The photon energy spectrum is defined as

\begin{equation}\label{enerspectr}
\frac{d\Gamma}{d E_\gamma}=
\frac{\alpha}{24\pi^2} \frac{1}{\sqrt{s_2}}E_\gamma^3 P_\pi^3
|G^{(0)}_{\eta\pi^+\pi^-\gamma}(s_2)|^2.
\end{equation}
with $s_2=m_P(m_P-2E_\gamma)$.

We will discuss the numerical results in the final subsection.

\subsection{Discussion of the numerical results}

The dependence of the ratios of the theoretical values of decay widths to 
their experimental averages on the mixing angle are shown in Fig. 3a 
Fig. 3b for the vector propagator shape described by Eq.~(\ref{propim_QCM}) 
and Eq.~(\ref{propim_VDM}), respectively. Visually, one can see that the value
of the mixing angle $\theta_P=-16.5^{\rm o}$ in Fig. 3a seems may give quite
resonable coincidence with the experimental data except the decay
$\eta\to\pi\pi\gamma$ which theoretical value is 1.5 times larger than
the average experimental one. One has to note that the value 
$\theta_P=-16.5^{\rm o}$ is supported by two recent
theoretical analysis: (1) $\eta,\eta^\prime\to\gamma\gamma^*$ 
and $D_s\to\eta l\nu/\eta^\prime l\nu$ which gives 
$\theta_P=-16.7^{\rm o}\pm 2.8^{\rm o}$ \cite{Anis}, and (2)
$J/\psi$ decays into a vector and pseudoscalar meson which gives
$\theta_P=-16.9^{\rm o}\pm 1.7^{\rm o}$ \cite{Bramon}.

First, we discuss the decays $\eta(\eta^\prime)\to\gamma\mu^+\mu^-$
and $\omega\to\pi^0\mu^+\mu^-$.  The transition form factors decsribed
the observable spectra Eqs.~(\ref{spec_p}-\ref{spec_v}) have been parametrized
in the pole approximation \cite{Land} as

\begin{eqnarray}
F_\eta(q^2)&=&\frac{1}{1-q^2/\Lambda^2_\eta} \hspace{1cm}
\Lambda_\eta=0.72\pm 0.09 \; {\rm GeV}, 
\label{lam_eta}\\
F_{\omega\pi}(q^2)&=&\frac{1}{1-q^2/\Lambda^2_\omega} \hspace{1cm}
\Lambda_\omega=0.65\pm 0.03 \; {\rm GeV}.
\label{lam_omega}
\end{eqnarray}

As  it is seen from Eqs.~(\ref{lam_eta}-\ref{lam_omega}) a value of
the characteristic mass, $\Lambda$, agrees well with vector dominance 
($m_\rho=0.77$ GeV) for $\eta$-decay, and differs from those by four
standard deviations for $\omega$-decay. Our form factor (see, Fig. 4) goes 
slightly lower than the curve fitted the experimental data but it is still 
within the experimental uncertainties. The behavior of
$\omega-\pi$-form factor (see, Fig. 5) almost coincides with the fitting curve.

The behavior of the form factor describing the $\eta^\prime\to\gamma\mu^+\mu^-$
decay is more interesting since the $\rho^0$ and $\omega$ poles occur
in the physical region allowed for the muon pair spectrum.
Experimental data on the $\eta^\prime$-meson transition form factor, the VDM
predictions with taking into account the $\rho$-meson decay width
in the vector propagators, and the results of our calculations are shown
in Fig. 6. A more detailed analysis is difficult to perform because of 
poor statistics. 

Now let us discuss the decay $\eta^\prime\to\gamma\pi^+\pi^-$ and related
process $\eta^\prime\to\rho^0\gamma$. There are lots of experiments
devoted to the study of these decays \cite{BNL1}-\cite{Lepton}.
Most of them are looking at the effective mass spectrum for the $\pi^+\pi^-$
system in the decay $\eta^\prime\to\pi^+\pi^-\gamma$ (see, Fig. 7).
It was shown in \cite{Lepton} that an attempt to fit the distribution
in the framework of the cascade decay model $\eta^\prime\to\rho^0\gamma$,
$\rho^0\to\pi^+\pi^-$ with simple choice of the vector propagator

$$
\frac{dN}{dm_{\pi\pi }}\propto \frac{E^3_\gamma P_\pi^3 m^3_{\pi\pi}}
{(m^2_{\pi\pi}-m^2_\rho)^2+m_\rho^2\Gamma_\rho^2}
$$
with $\Gamma_\rho$ being the "dynamical" width of $\rho$-meson is not 
successful because of the distribution obtained in experiment is much harder
than the theoretical one. It was needed to assume that either
the propagator form should be modified significantly without resonable
theoretical justifications or a nonresonance process gives significant
contribution to this decay. The last assumption was realized \cite{Lepton}
by modelling the distribution by the following form

$$
\frac{dN}{dm_{\pi\pi }} \propto E^3_\gamma P_\pi^3 m^3_{\pi\pi}
|\frac{1}{ m^2_{\pi\pi}-m^2_\rho+im_\rho\Gamma_\rho}+
\frac{\xi e^i\alpha}{m^2_{\eta^\prime}}|^2.
$$
As a result of fitting the experimental spectrum it was found that
$\xi=2.78\pm 0.46$ and $\alpha=-1.07\pm0.08$. 

Recent measurements of the $\pi^+\pi^-$ mass spectrum in the decay
$\eta^\prime\to\pi^+\pi^-\gamma$ with the Crystal Barrel detector 
\cite{crystal}
also confirmed the existence of a non-resonant coupling $\pi^+\pi^-\gamma$.
We plot their data on Fig. 7a and compare them with our results and
VDM prediction. It is readily seen that our results coincide quite well with
the experimental data whereas the VDM predictions are significantly above them
up to the resonant region.

The existence of the non-resonant
contribution clearly comes from the existence of the box anomaly in
chiral theory. One of the attempts to incorporate vector mesons in the chiral
Lagrangian has been done in \cite{Picc}. It was found that the amplitude
of the decay $\eta\to\pi^+\pi^-\gamma$ is given by

\begin{equation}\label{Chir}
M(\eta\to\pi^+\pi^-\gamma)=\frac{e}{4\pi^2 F_\pi^2}
\epsilon^{\mu\nu\alpha\beta} p^+_\mu p^-_\nu q_\alpha \varepsilon_\beta
\biggl[
\frac{1}{2}-\frac{3}{2} \frac{m^2_\rho}{m^2_\rho-p^2_\rho}
\biggr]
[
\frac{1}{\sqrt{3}f_8} \cos\theta-\sqrt{\frac{2}{3}} \frac{1}{f_0}\sin\theta
]
\end{equation}
where $f_8=1.25F_\pi$ and $f_0=1.04F_\pi$. A numerical result for the decay
width obtained with $\theta=-20.6$ was $\Gamma(\eta\to\pi^+\pi^-\gamma)$= 62 eV
which is close to the experiment $\Gamma_{\rm expt}=58\pm 6$ eV.
The direct generalization of this model for $\eta^\prime$-decay
by adding the imaginary part like $im_\rho\Gamma_\rho$ to the denominator
of propagator and replacing $\cos\theta$ to $\sin\theta$ and
$\sin\theta$ to -$\cos\theta$, respectively, gives the spectrum
and the decay width listed in Fig. 7b in a agreement with the experimental
data and  our results.

Now we  discuss the $\eta\to\pi^+\pi^-\gamma$-decay. From the theoretical point
of view this is simpler than the $\eta^\prime$-decay since it is up to
the resonance region. But experimental data are very poor, actually,
there are only two experimental papers discussing this issue \cite{Layter}
and \cite{Gormley} where the $\gamma$-ray energy spectrum was measured.
To compare the theoretical results with the experimental data,
two forms of function $f(s)$ entering to the matrix element
$M(\eta\to\pi^+\pi^-\gamma)\propto f(s)\varepsilon^{\mu\nu\alpha\beta}
\epsilon^\mu_\gamma p_\gamma^\nu p_+^\alpha p_-^\beta$
have been used \cite{Layter}: $f(s)=1$ and $f(s)=m^2_\rho/(m^2_\rho-s)$.
The efficiency used in this procedure has not been given in explicit form.
We are able to reproduce the results of the paper \cite{Layter}
with the efficiency function chosen in the form

\begin{equation}\label{eff}
{\rm eff}(E_\gamma)=E_\gamma^a (E_\gamma^{(0)}-E_\gamma)^b
\end{equation}
where $a=b=1.1$. Using this efficiency function we recalculate the
experimental data according to 
$$
{\rm corrected \;\;data}={\rm data}/{\rm eff}(E_\gamma) 
$$
and put them in Fig. 7a and Fig. 7b. One can see that the theoretical results
from QCM, VDM, and the chiral approach implemented VDM \cite{Picc}
coincide each other and fit very well the experimental data.
Finally we summarize our predictions for the rates of radiative decays of 
$\eta$ and $\eta^\prime$ mesons in Table III.

\begin{center}
{\bf ACKNOWLEDGMENTS}
\end{center}

This work was supported in part by the United States Department of Energy
under Grant No. DE-FG-ER40413.

\newpage
\normalsize

\begin{center}
{Table I. Variuos functions entering the QCM calculations
of meson vertecies. Functions $a(u)$ and $b(u)$ characterize
the confined quark propagator Eq.~(\ref{conf}-\ref{confunc}).\\}
\end{center}
\large
\begin{center}
\def\arraystretch{2.0}
\begin{tabular}{|l|}
\hline\hline
$R_{PP}(x)=B_0+\frac{x}{4}\int\limits_0^1dub(-\frac{xu}{4})
\frac{(1-u/2)}{\sqrt{1-u}}$\\
\hline
$R_{VV}(x)=B_0+\frac{x}{4}\int\limits_0^1dub(-\frac{xu}{4})
\frac{(1-u/2+u^2/4)}{\sqrt{1-u}}$\\
\hline
$R_{V}(x)=B_0+\frac{x}{4}\int\limits_0^1dub(-\frac{xu}{4})
(1+u/2)\sqrt{1-u}$\\
\hline
$R_{VPP}(x)=B_0+\frac{x}{4}\int\limits_0^1dub(-\frac{xu}{4})\sqrt{1-u}$\\
\hline
$R_{PVV}(x,y,z)=\int d^3\alpha \delta(1-\sum\limits_1^3\alpha_i)
a(-\alpha_1\alpha_2 x-\alpha_1\alpha_3 y-\alpha_2\alpha_3 z)$\\
$R_{PVV}(x,y,0)=(xR_{PVV}(x,0,0)-yR_{PVV}(y,0,0))/(x-y)$ \\
$R_{PVV}(x,0,0)=\frac{1}{4}\int\limits_0^1du a(-\frac{xu}{4})
\ln\biggl({\frac{1+\sqrt{1-u}}{1-\sqrt{1-u}}}\biggr)$\\
\hline
\hline
$A_0=\int\limits_0^\infty dua(u)=1.09$ \hspace{1cm}
$B_0=\int\limits_0^\infty dub(u)=2.26$ \\
\hline\hline
\end{tabular}
\end{center}

\newpage
\normalsize

\begin{center}
{Table II. The SU(3)-factors entering the calculations of
$P\to\gamma\gamma$ ($P\to\gamma l^+l^-$); \\
$V\to P\gamma$ ($V\to P l^+l^-$), $P\to V\gamma$; and 
$V\to\gamma\to l^+l^-$ decays. \\}
\end{center}
\begin{center}
\def\arraystretch{2.0}
\begin{tabular}{|c|c|}
\hline\hline
$C_{P\gamma\gamma}=3{\rm tr}(\lambda^P Q^2)$   & P \\
\hline\hline
 1                                         & $\pi^0$ \\
\hline
-$(5\sin\delta_P+\sqrt{2}\cos\delta_P)/3$ &  $\eta$         \\
\hline
$(5\cos\delta_P-\sqrt{2}\sin\delta_P)/3$ &  $\eta^\prime$ \\
\hline\hline
\end{tabular}
\end{center}

\begin{center}
\def\arraystretch{2.0}
\begin{tabular}{|c|c|}
\hline\hline
$C_{VP\gamma}={\rm tr}(Q\{\lambda^P,\lambda^V\})$ & $VP\gamma$ \\
\hline\hline
2/3 & $\rho\pi\gamma$ \\
\hline
2 &  $\omega\pi\gamma$ \\
\hline
$-2\sin\delta_P$ &  $\rho\eta\gamma$ \\
\hline
$-(2/3)\sin\delta_P$ &  $\omega\eta\gamma$ \\
\hline
$-(4/3)\cos\delta_P$ &  $\phi\eta\gamma$ \\
\hline
$2\cos\delta_P$ &  $\rho\eta^\prime\gamma$ \\
\hline
$(2/3)\cos\delta_P$ &  $\omega\eta^\prime\gamma$ \\
\hline
$-(4/3)\sin\delta_P$ &  $\phi\eta^\prime\gamma$ \\
\hline\hline
\end{tabular}
\end{center}

\begin{center}
\def\arraystretch{2.0}
\begin{tabular}{|c|c|}
\hline\hline
$C_{V\gamma}={\rm tr}(\lambda^V Q)$   & V \\
\hline\hline
 1             &  $\rho^0$ \\
\hline
1/3            &  $\omega^0$         \\
\hline
$\sqrt{2}/3$   &  $\phi$ \\
\hline\hline
\end{tabular}
\end{center}

\newpage
\normalsize

\begin{center}
{Table III. Comparison of our predictions for radiative decay widths
of neutral pseudoscalar and vector mesons with the available
experimental data. Results are given for the 
$\eta-\eta^\prime$-mixing angle $\theta_P=-16.5^{\rm o}$.\\}
\end{center}

\large
\begin{center}
\def\arraystretch{2.0}
\begin{tabular}{|l|c|c|}
\hline\hline
     & {\rm QCM} & {\rm Expt.} \\
\hline\hline
$\pi^0\to\gamma\gamma, \;\; {\rm eV}$ & 7.2 & $7.7\pm 0.6$ \\
\hline
$\eta\to\gamma\gamma, \;\; {\rm KeV}$        & 0.45   & $0.46\pm 0.04$ \\
\hline
$\eta^\prime\to\gamma\gamma, \;\; {\rm KeV}$ & 4.2    & $4.26\pm 0.19$ \\
\hline
$\rho^0\to\eta\gamma, \;\; {\rm KeV}$        & 38     & $57\pm 12$ \\
\hline
$\eta\to\mu^+\mu^-\gamma, \;\; {\rm eV}$ &  0.35 & $0.37\pm 0.06$ \\
\hline
$\eta\to\pi^+\pi^-\gamma, \;\; {\rm eV}$ &  92 & $58\pm 6$ \\
\hline
$\eta^\prime\to\mu^+\mu^-\gamma, \;\; {\rm KeV}$ & 0.014 & $0.021\pm 0.007$ \\
\hline
$\eta^\prime\to\pi^+\pi^-\gamma, \;\; {\rm KeV}$ & 47 & $56\pm 9$ \\
\hline
$\eta^\prime\to\rho^0\gamma, \;\; {\rm KeV}    $ & 38  &  $61\pm 8$\\
\hline
$\omega\to\pi\gamma, \;\; {\rm KeV}$ & 590 & $716\pm 51$ \\
\hline
$\omega\to\pi\mu^+\mu^-, \;\; {\rm KeV}$ & 0.49 & $0.81\pm 21$ \\
\hline\hline
\end{tabular}
\end{center}

\newpage
\normalsize
\baselineskip 20pt

\noindent
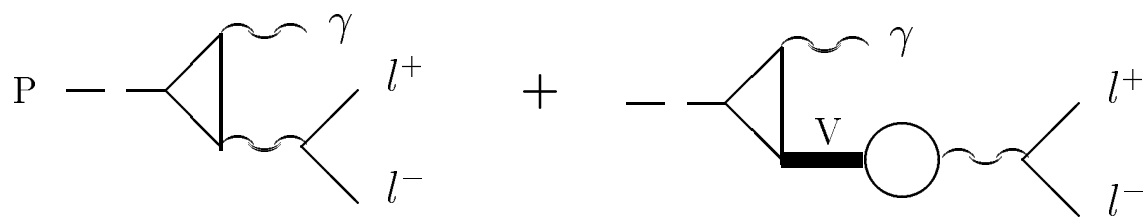
Fig. 1. The $P\to\gamma\l^+l^-$ decay.\\
\noindent
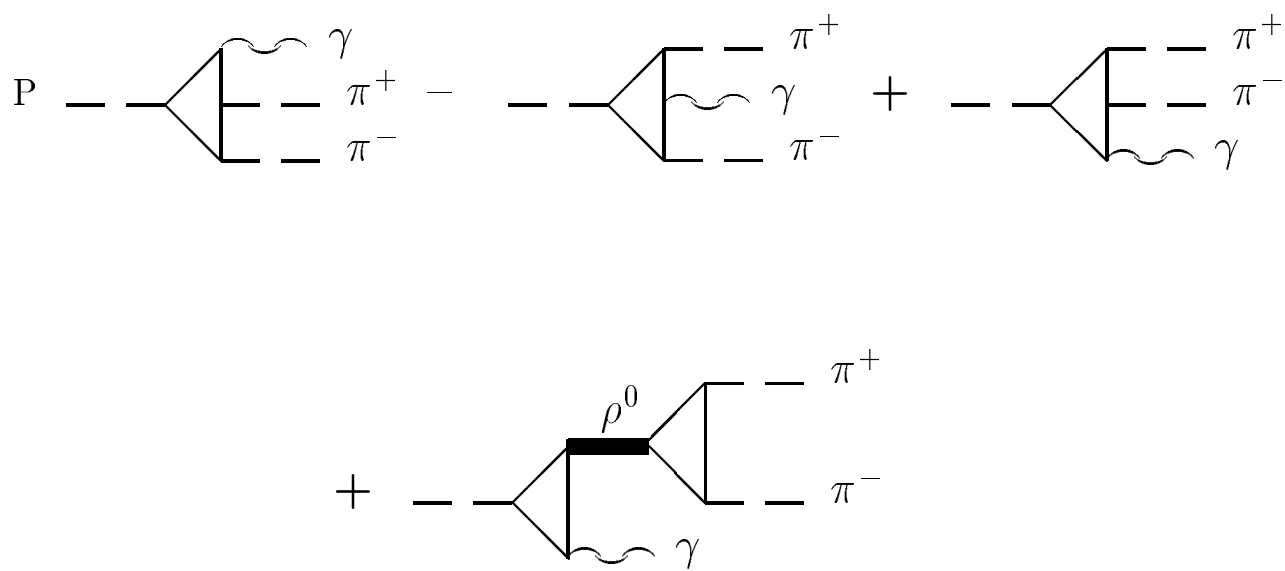
Fig. 2. The $P\to\pi^+\pi^-\gamma$ decay.\\
\noindent
Fig. 3. The dependence of ratios of theoretical values for the radiative decays
of $\eta$ and $\eta^\prime$ mesons to their experimental averages on
the mixing angles calculated with  the shape of vector meson propagators 
defined by (a) Eq.~(\ref{propim_QCM}), and (b)  Eq.~(\ref{propim_VDM}).  \\
\noindent
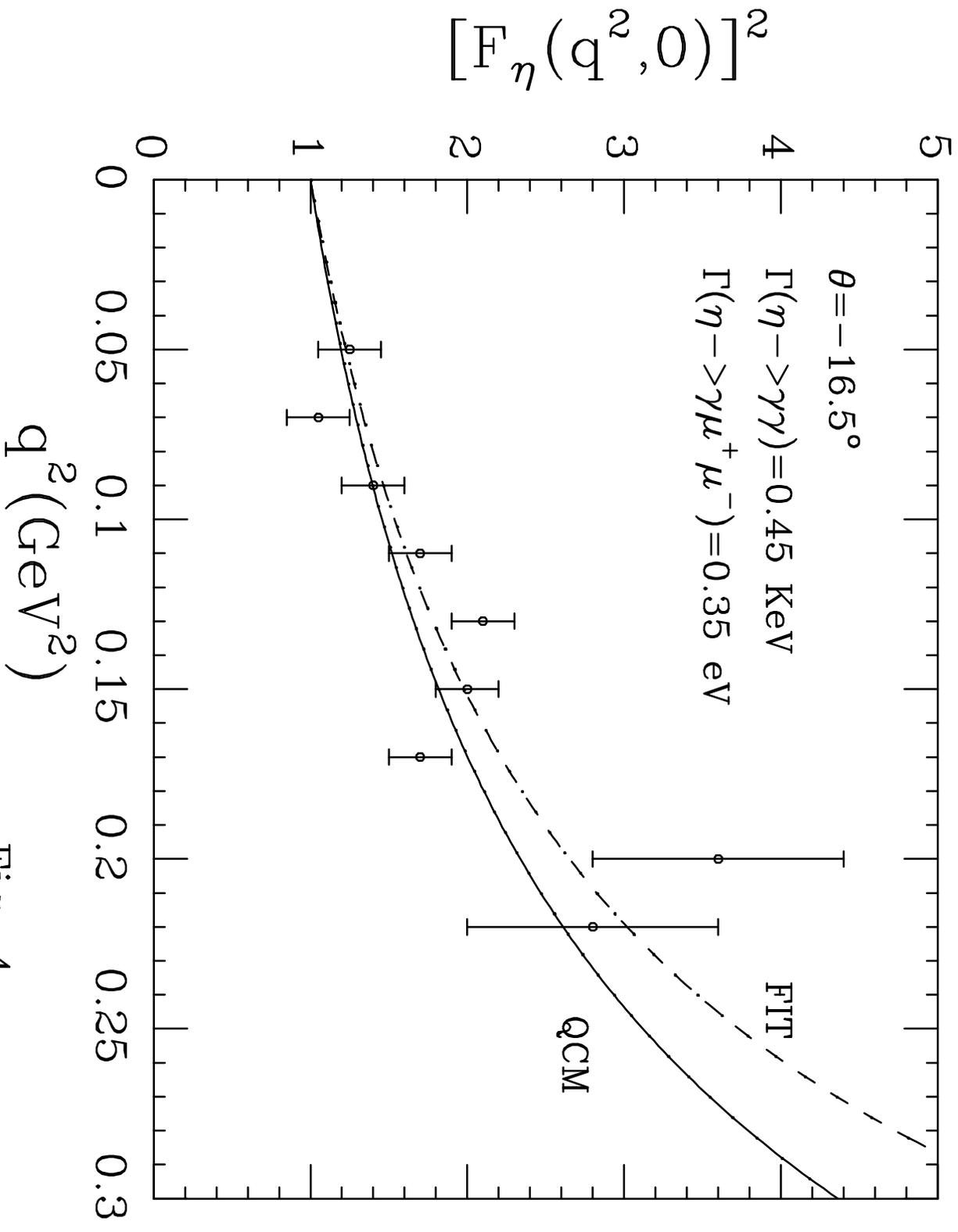
Fig. 4. The form factor of $\eta\to\gamma\mu^+\mu^-$ decay.\\
\noindent
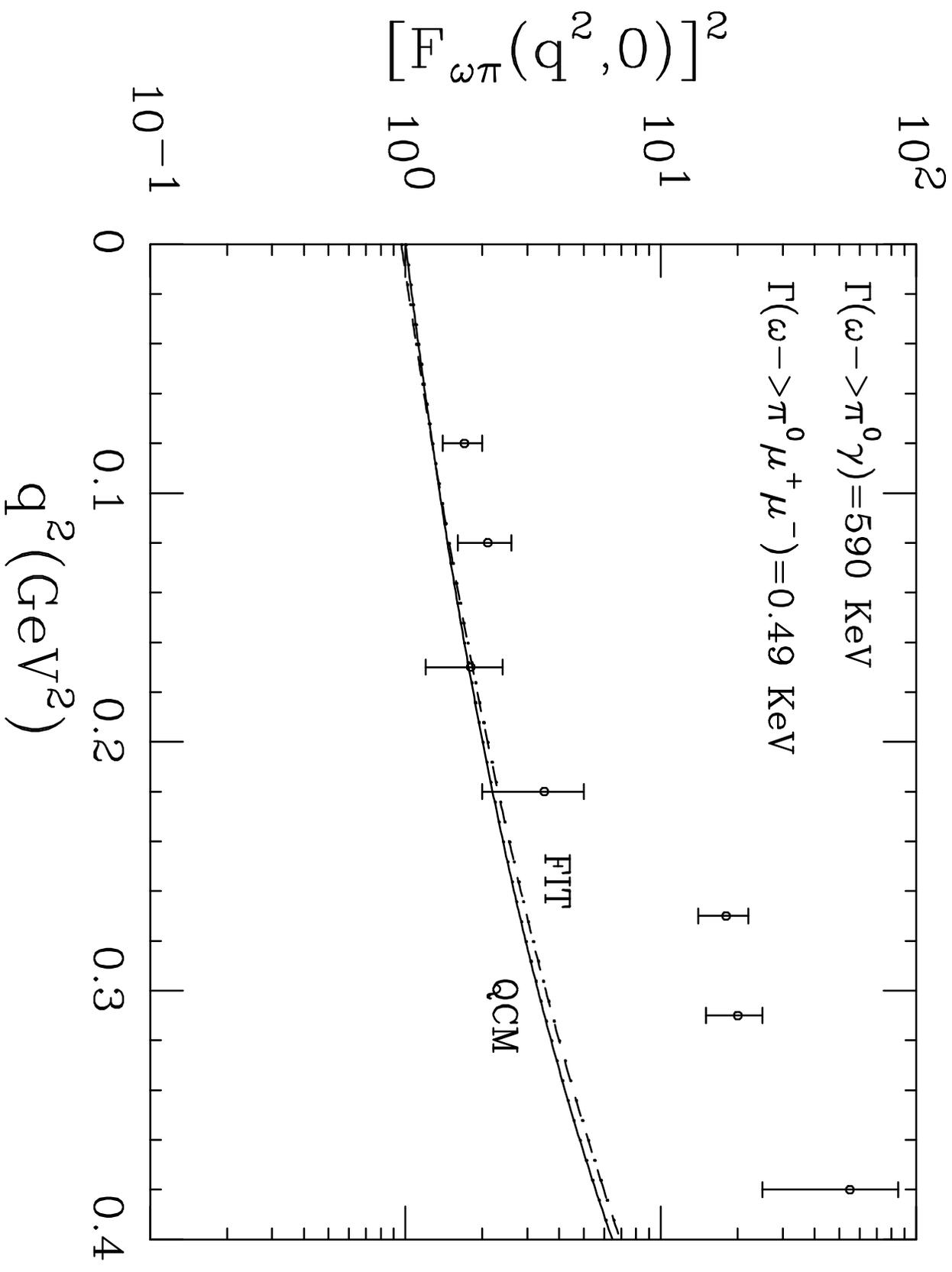
Fig. 5. The form factor of $\omega\to\pi^0\mu^+\mu^-$ decay.\\
\noindent
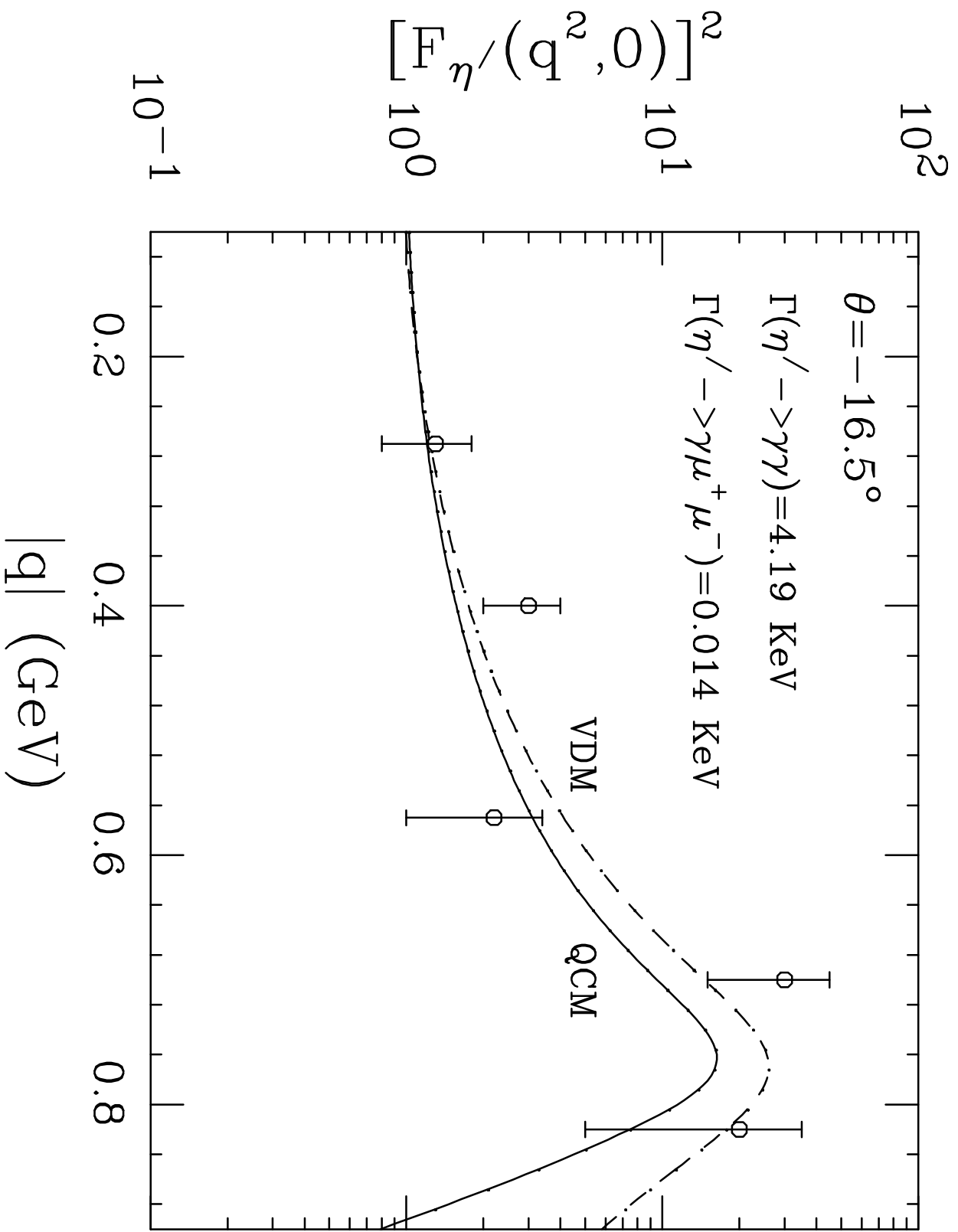
Fig. 6. The form factor of $\eta^\prime\to\gamma\mu^+\mu^-$ decay.  \\
\noindent
Fig. 7. The mass spectrum of $\eta^\prime\to\gamma\pi^+\pi^-$ decay:\\
(a) QCM and VDM (b) Chiral model \cite{Picc} and VDM.\\
\noindent
Fig. 8. The photon energy spectrum of $\eta\to\gamma\pi^+\pi^-$ decay:\\
(a) QCM and VDM, (b) Chiral model \cite{Picc} and VDM.\\
 
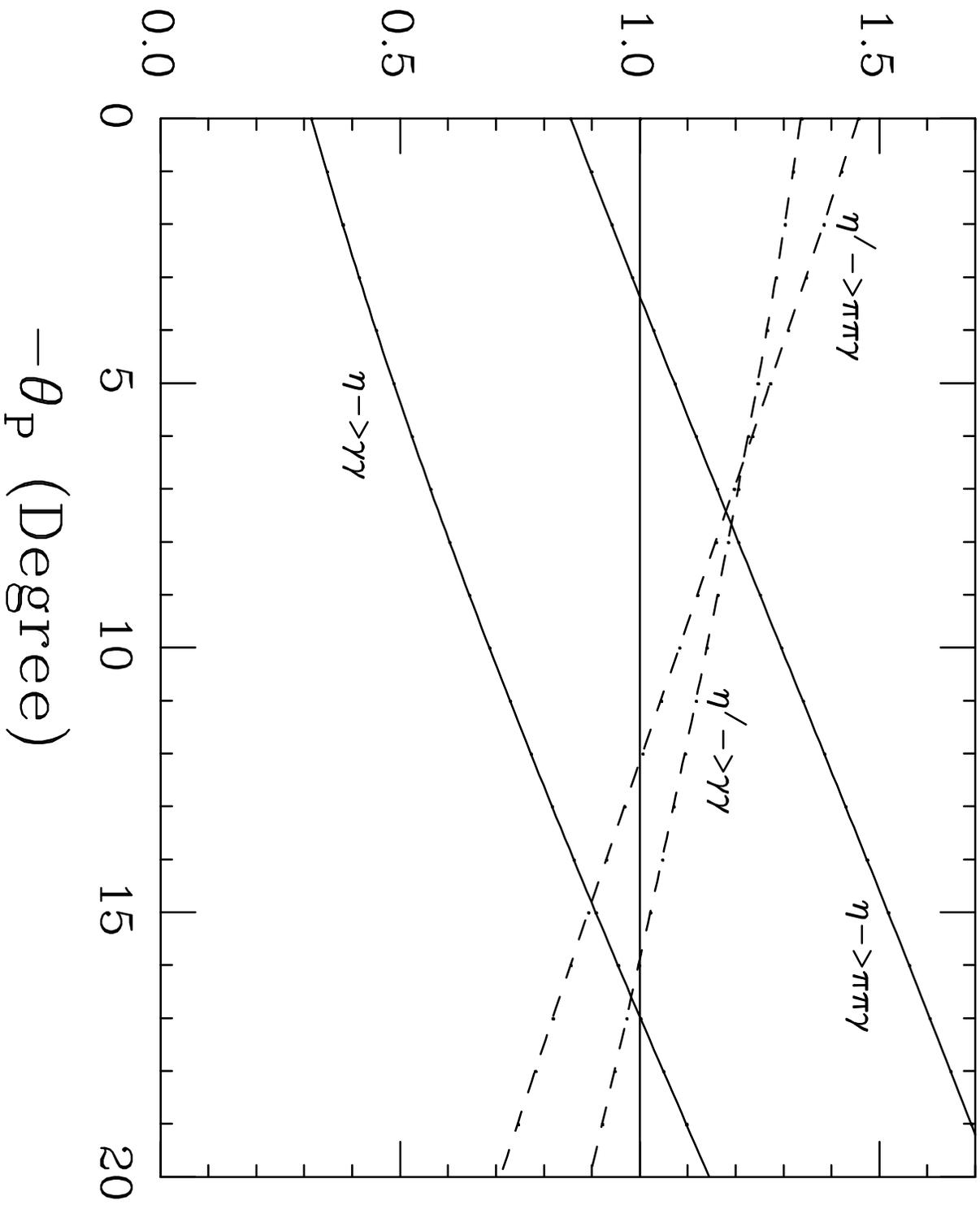
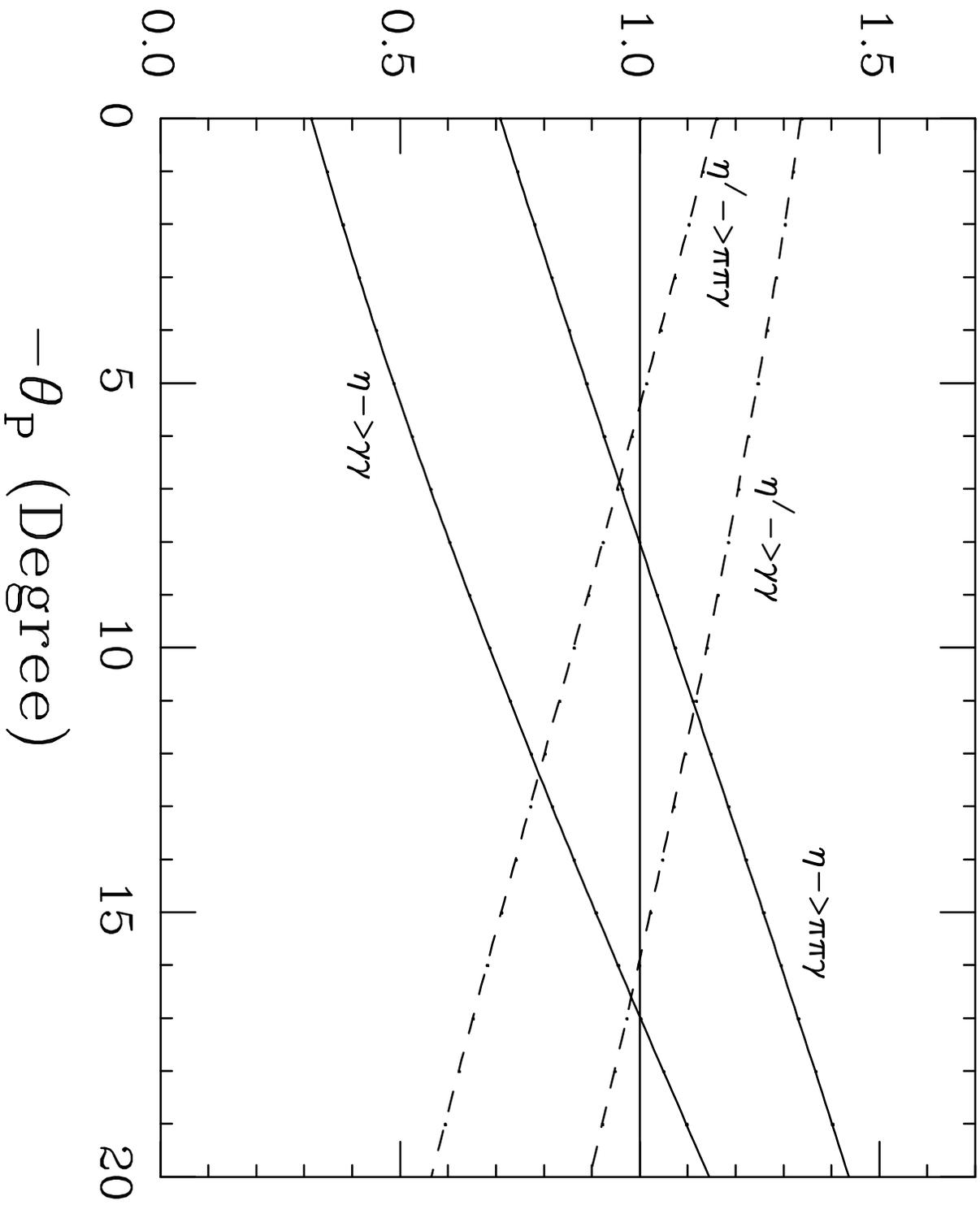
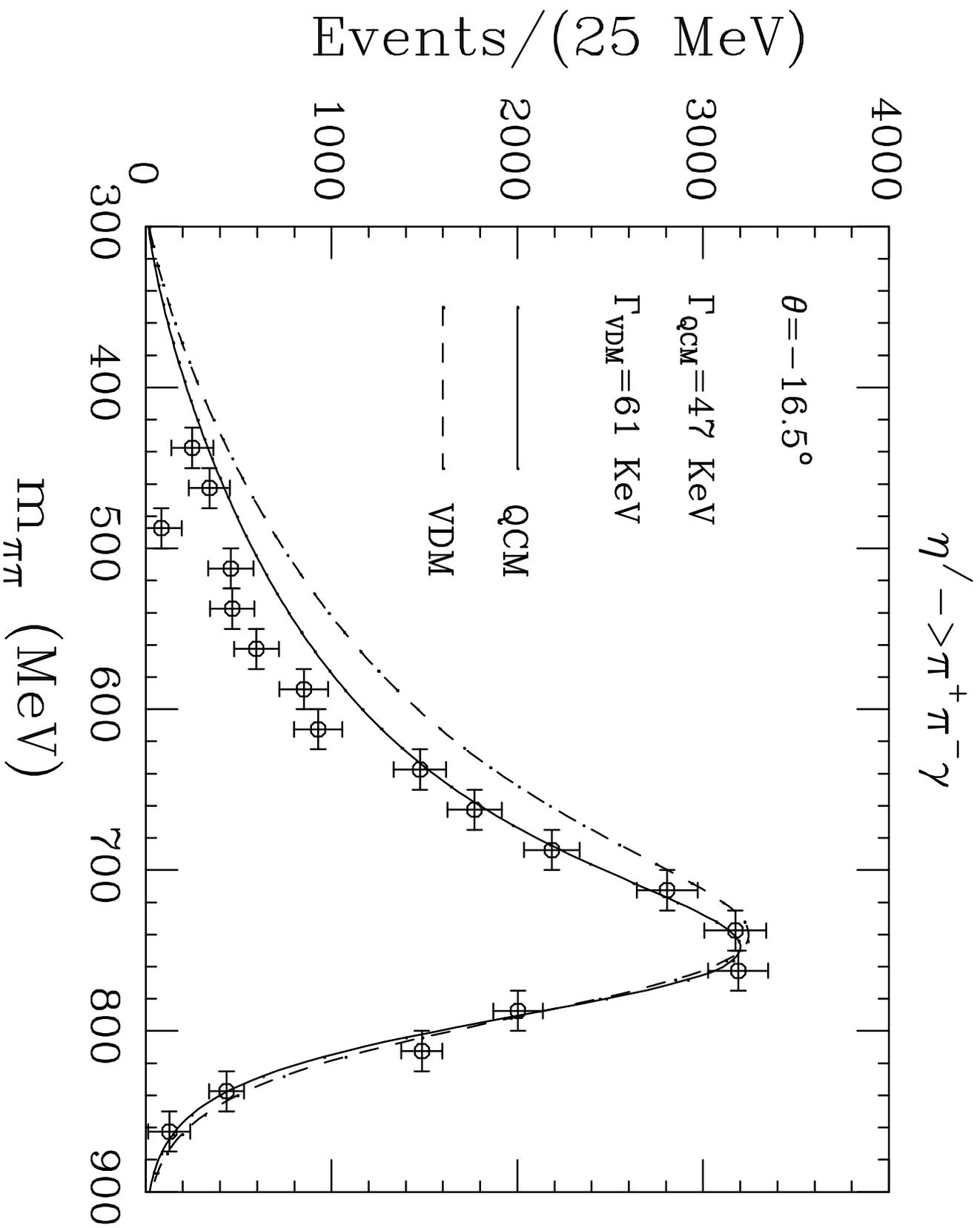
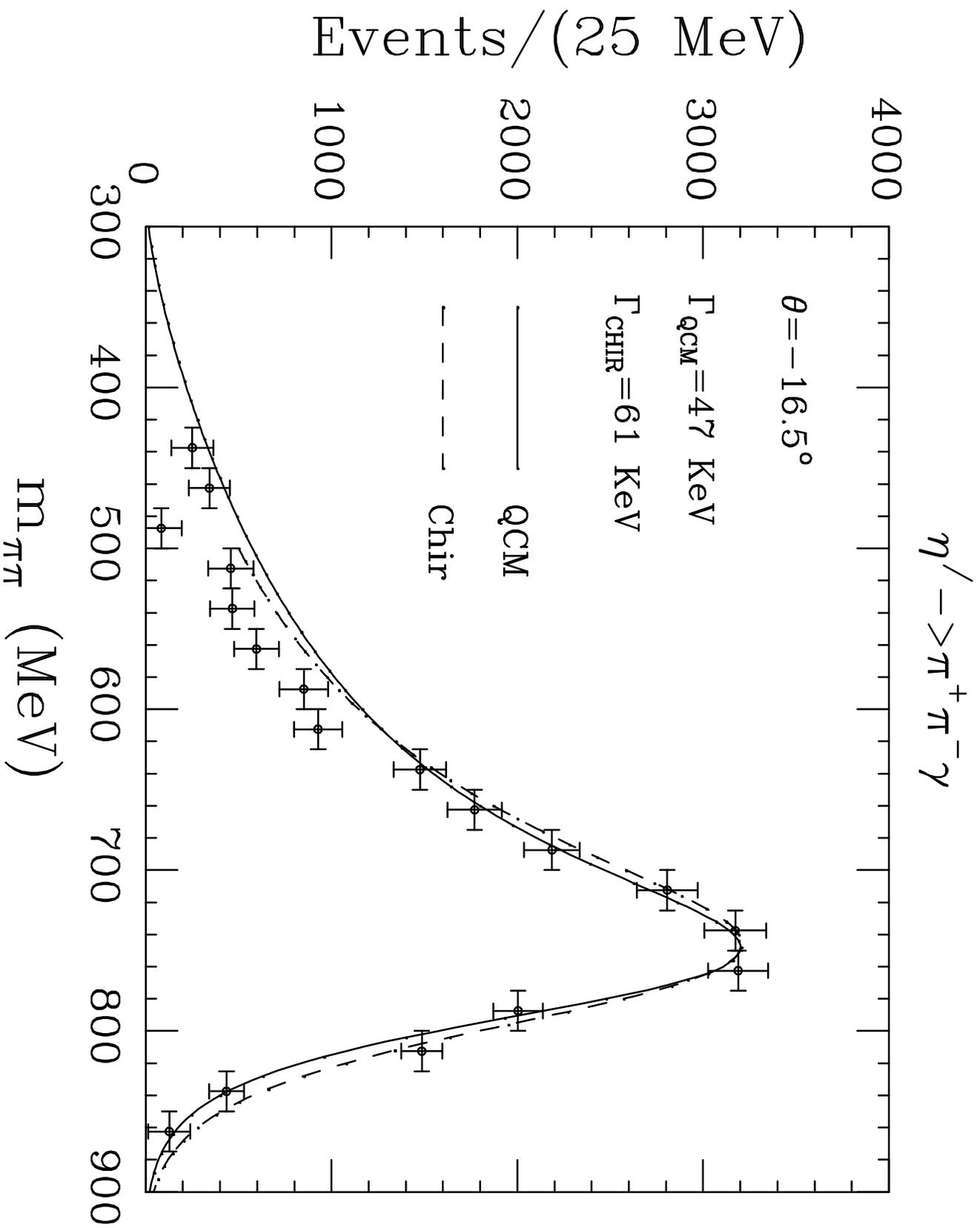
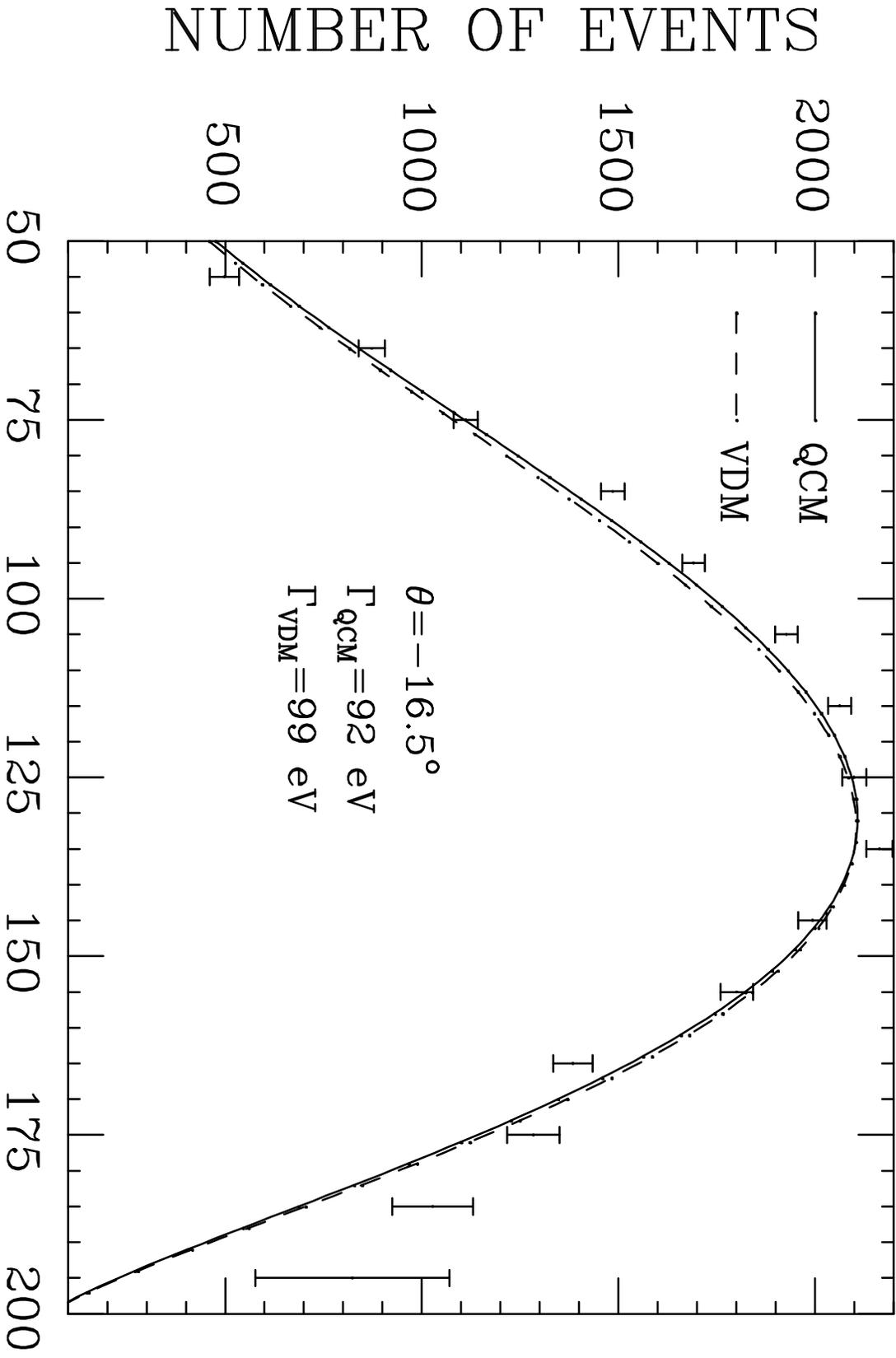
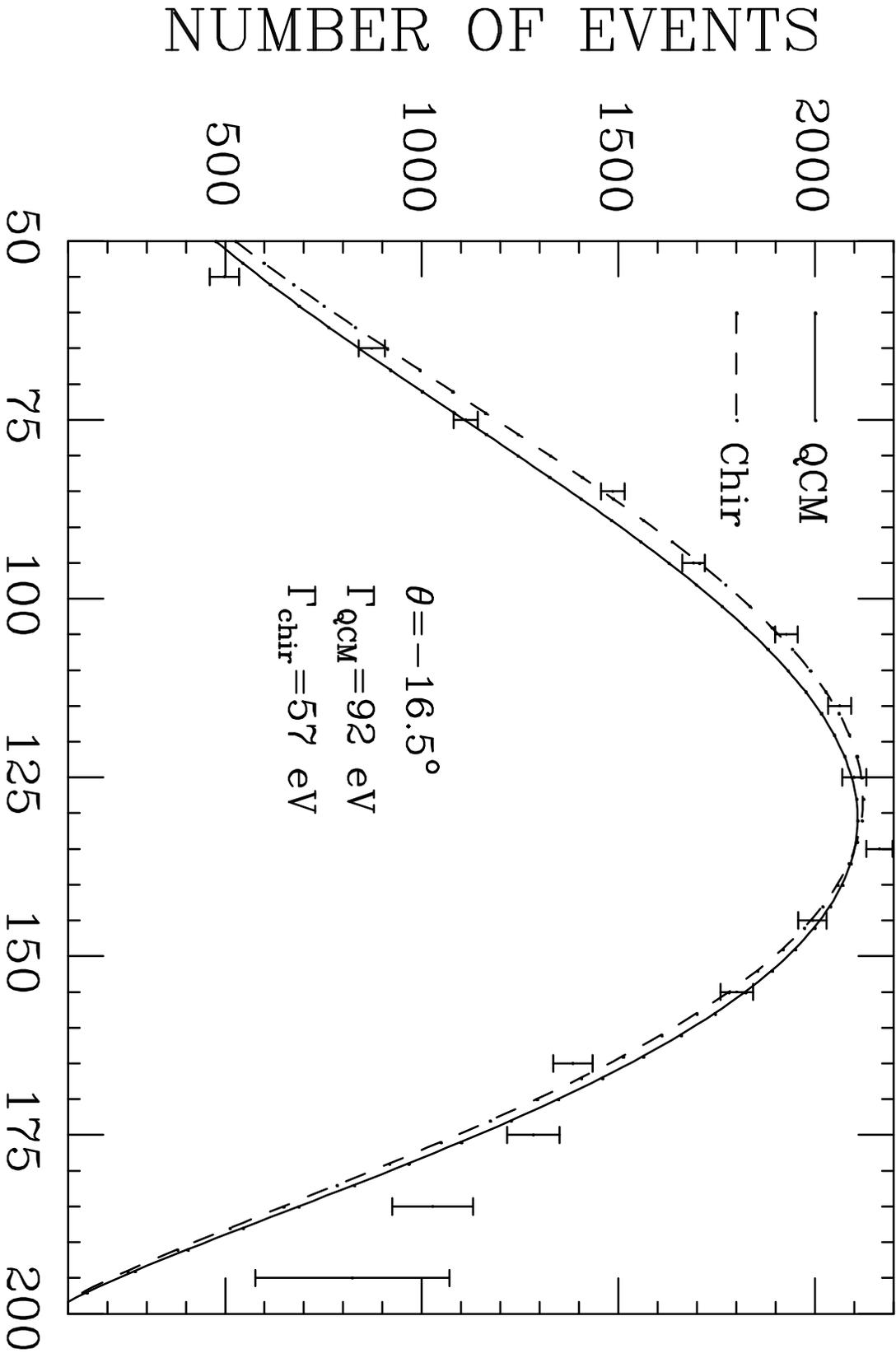
\end{document}